# Nested-capillary anti-resonant silica fiber with mid-infrared transmission and low bending sensitivity at 4000 nm

Mariusz Klimczak,[1,3,*] Dominik Dobrakowski,[1,2] Amar Nath Ghosh,[3] Grzegorz Stępniewski,[1,2] Dariusz Pysz,[1] Guillaume Huss,[4] Thibaut Sylvestre,[3] Ryszard Buczyński[1,2]

[1]*Institute of Electronic Materials Technology, Glass Department, Wólczyńska 133, 01-919 Warsaw, Poland*
[2]*University of Warsaw, Faculty of Physics, Pasteura 5, 02-093 Warsaw, Poland*
[3]*Institut FEMTO-ST, CNRS, UMR 6174, Université Bourgogne Franche-Comté, Besançon, France*
[4]*LEUKOS, 37 rue Henri Giffard, Z.I. Nord, 87280 Limoges, France*
*\*Corresponding author: [mariusz.klimczak@itme.edu.pl](mariusz.klimczak@itme.edu.pl)*

**We report a silica glass nested capillary anti-resonant nodeless fiber with transmission and low bending sensitivity in the mid-infrared around 4000 nm. The fiber is characterized in terms of transmission over 1700-4200 nm wavelengths, revealing a mid-infrared 3500-4200 nm transmission window, clearly observable for a 12 m long fiber. Bending loss around 4000 nm is 0.5 dB/m measured over 3 full turns with 40 mm radius, going up to 5 dB/m for full turns with 15 mm radius. Our results provide experimental evidence of hollow-core silica fibers in which nested, anti-resonant capillaries provide high bend resistance in the mid-infrared. This is obtained for a fiber with large core diameter of over 60 µm relative to around 30 µm-capillaries in the cladding, which motivates its application in gas fiber lasers or fiber-based mid-infrared spectroscopy of $CO_x$ or $N_xO$ analytes.**

Hollow core glass fibers (HCFs) relay on guiding mechanisms, which enable significant reduction of overlap of the guided mode with solid microstructure forming the cladding. The inhibited coupling fibers (ICFs) and hollow core anti-resonant guiding fibers (ARFs) are the two examples, which received particularly strong attention. The first reported ICFs were the Kagomé fibers, developed in 2002 [1]. The principle of light guidance in ICFs was explained in 2007 [2], revealing the fundamental difference between photonic bandgap (PBG) and inhibited coupling (IC). ARFs can be considered as simplified ICF structures, consisting of only one ring of circular capillaries in the cladding. These fibers were first reported in 2011 [3] and the anti-resonant reflecting optical waveguide (ARROW) model [4] is considered the most accurate in describing guiding in these structures. Various designs of hypocycloid-core fibers have followed since, with emphasis on fibers with the core area limited by a single ring of circular, non-touching capillaries [5,6]. Low intrinsic nonlinearity and dispersion if these fibers make them particularly attractive for high energy pulse delivery [7]. The possibility to largely modify their optical properties by infiltration with liquids or gases opens interesting applications in optofluidics [8] or in temporal compression down to single optical cycles of laser pulses in the mid-infrared [9]. Low attenuation is the obvious advantage of any fiber for a practical application and loss below 10 dB/km in the visible and at important laser wavelengths in the near-infrared has been reported in HCFs [10,11]. More recently, an ARF with measured attenuation of 2 dB/km at a wavelength of 1512 nm made it possible to consider them for specific telecommunication applications [12]. Such applications are among the important motivations for development of HCFs and data transmission was demonstrated in an air-core PBG fiber already in 2013 [13]. Dramatic improvement in bringing down of attenuation – ultimately to around 1 dB/km – prompted later demonstrations of data transmission in hollow core ARFs in the third telecommunication window, as well [14-16]. Mid-infrared guidance in silica fibers is disruptive for gas sensing and spectroscopy applications [5,17]. The feasibility to design robust multiple-pass setups is crucial for development of fiber-based cavity-enhanced spectroscopy [18]. Several-meter or longer gas-filled fibers are becoming important for fiber-based gas lasers exploiting Raman scattering [19]. More recently, mid-infrared lasers operating between 3100 nm or 4600 nm wavelengths and based on optical transitions between vibrational energy bands of different molecular gasses have been reported using gas-filled HCFs [20-23]. The fiber cavity length, exceeding 10 m in either of cases, was reported among the key parameters of optimizing laser efficiency, which puts significant pressure on bending loss performance of fibers used to build such systems. Gas-filled HCFs have also been shown as attractive experimental platforms for nonlinear propagation, broadening and compression of ultrashort pulses across the UV to the mid-infrared [24-27]. Silica glass ARFs have been experimentally demonstrated to guide light at wavelengths up to 7900 nm, with attenuation at 3900 nm of roughly 50 dB/km [28]. Attenuation of 34 dB/km at 3050 nm and 85 dB/km at 4000 nm, was found for ARFs with a triangular cladding [29,30].

Mode confinement in HCFs is very sensitive to bending [31]. Hollow core PBG fibers were outperforming ARFs in this regard [31]. Significant research effort has been devoted to overcoming this drawback, resulting in reported bend losses of ICFs in the order of 0.25 dB/turn [32], 0.2 dB/m of attenuation at 5 cm radius in the NIR range [33] and finally a result of 0.03 dB/turn bend loss for a 30 cm bend diameter at 750 nm has been reported [12]. Bend loss below 1 dB/km for R=10 cm at 1512 nm has recently been demonstrated as well [12]. An important conclusion has been drawn between low bending loss and small core diameter ARFs [34]. A variant of ARFs with nested capillaries, later commonly referred to as nested capillary anti-resonant nodeless fiber (NANF), has been shown to alleviate the challenge of bend losses [35]. The first report on a physical NANF structure involved a 10-capillary fiber, followed by a 5-capillary fiber [36,37].

In this Letter, we report on development of a silica glass NANF, and on characterization of its transmission, attenuation and bending losses with a focus on the mid-infrared spectral range around wavelength of 4000 nm.

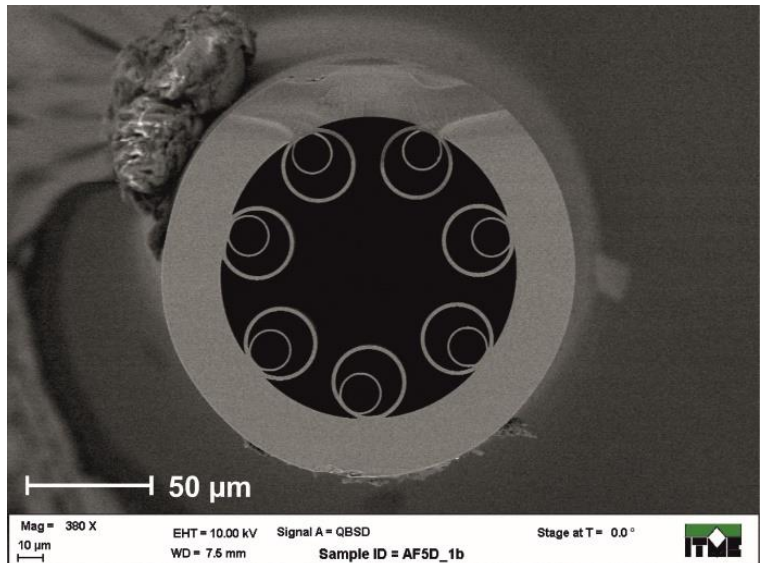

Fig. 1. SEM image of the developed fiber.

Our silica glass NANF was fabricated using the stack-and-draw technique. Microstructure of the fiber is shown in a scanning electron microscopy (SEM) image in Fig. 1. It consists of a cladding with 7 nested capillaries. Criterion for number of capillaries is usually attenuation of fiber and exceptionally low attenuation has been reported for ARFs with between 5 [15,38], through 6-7 [10,31] to 8-10 capillaries [5] in the cladding. A 7-nested capillary ARF has been reported previously, albeit with significantly less uniform structure, than that in Fig. 1 [39] and NANFs have been numerically investigated for mid-infrared transmission [40]. The outer diameter of our fiber is 162 µm, the air core diameter is 62 µm. The outer and inner capillary diameters are 29 µm and 16 µm, and their wall thickness is 1.6 µm and 0.9 µm, respectively.

All transmission measurements and mode imaging has been performed with a 150 cm long sample. Light from a Leukos Electro-250-Mir supercontinuum or NKT Photonics SuperK MIR supercontinuum has been coupled to the fiber with a mid-infrared, black diamond aspheric lens (Thorlabs C036TME). At output of the fiber, light was collimated with a parabolic off-axis silver mirror with a 10 cm focal length into a free-space input of a Fourier transform infrared (Thorlabs OSA205, FTIR) optical spectrum analyzer (OSA) for measurements in the mid-infrared or to a diffractive OSA for long-wave near-infrared measurements (Yokogawa AQ6375B). This, together with securing of both fiber ends in 9 cm long bare fiber mounts on translation stages, assured mechanical stability during measurements. In a similar way, the fiber output was collimated for characterization using an extended InGaAs beam profiler.

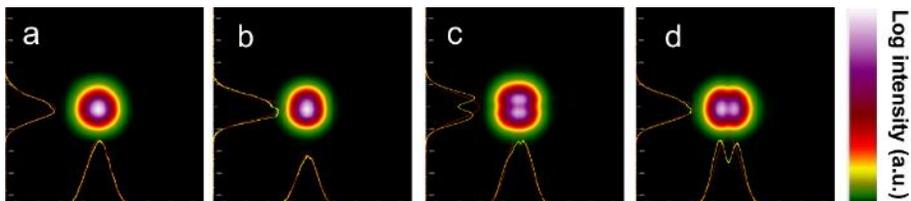

Fig. 2. Beam profiler images of fiber output recorded at 2500 nm: (a) fundamental mode (FM) in effectively straight fiber, (b) FM in fiber coiled at 0.5 cm radius, (c) and (d) $LP_{11}$ modes with the fiber effectively straight and coiled at 0.5 cm radius, respectively.

The fiber supports guidance of few modes at 2500 nm, but the fundamental mode can be selectively excited, as shown in Fig. 2(a), where light from an optical parametric oscillator has been used with around 100 nm of spectral width. Higher order modes in this case were excited using offset launch, as shown in Fig. 2(c). We also measured numerical aperture (NA) of the fiber by registering output beam profiles in far field with a phosphor-covered CCD camera. Output of single-mode laser diode operating at 1548 nm was coupled into the fiber with an aspheric lens (NA=0.25). Fundamental mode or higher order modes were selectively excited by changing the coupling conditions. Output beam size was measured at $1/e^2$ of maximum intensity. NA was calculated from the slope of linear fitting of measurement points (beam radius as a distance function) using formulas $slope = tg(\varphi)$ and $NA = sin(\varphi)$, obtaining NA = 0.037 for the fundamental mode and NA = 0.077 for a higher order mode. The fiber was protected with an acrylic coating and bending as tight as to a radius of R = 0.5 cm was possible. Depending on coupling adjustment, the $LP_{11}$ mode could be observed, as shown in Figs. 2(c,d). Bending of fiber with these modes excited resulted in scrambling to a structure similar to the fundamental mode.

Fig. 3 contains results related to transmission and results of attenuation measurements are shown in Fig. 4. Finite element method simulation results, obtained using Comsol Multiphysics (Wave optics module) for the fiber based on a vectorized image of real fiber structure, are shown in Fig. 3(a). Triangular mesh was used with size from 0.4 µm to 0.015 µm over glass parts of structure and between 0.6 µm and 0.06 µm over the air core. A 3 µm thick perfectly matched layer was assumed for the boundary condition. Material loss of silica was neglected. Bending loss was modelled assuming straight fiber geometry with modified refractive index [41]. Simulation results predict transmission windows in the 1800-2600 nm spectral range and over 3400-5000 nm. Attenuation penalty of around a factor of 10 at a wavelength of about 4500 nm can be anticipated, and this is related to the fact, that equal capillary thickness has not been maintained in the technological process. Numerical simulations of bend loss further reveal, that confinement loss penalty of about 1 dB/m can be expected, regardless of bend radius, when comparing the developed fiber to an equal capillary thickness (0.9 µm) structure. Up to the wavelength of around 4200 nm there is good agreement between the simulated and measured transmission spectra, as shown in Fig. 3(b). Compared to single capillary ring structure, simulations show that the nested capillary design would have a magnitude lower bend loss in each of the transmission windows, and similarly lower straight fiber attenuation, although straight fiber advantage would be lost closer to 5000 nm due to uneven capillary wall thickness. Simulations allow to expect transmission in the mid-infrared window to extend at least to 5000 nm, although we did not have light sources with spectral coverage above 4200 nm to verify this experimentally.

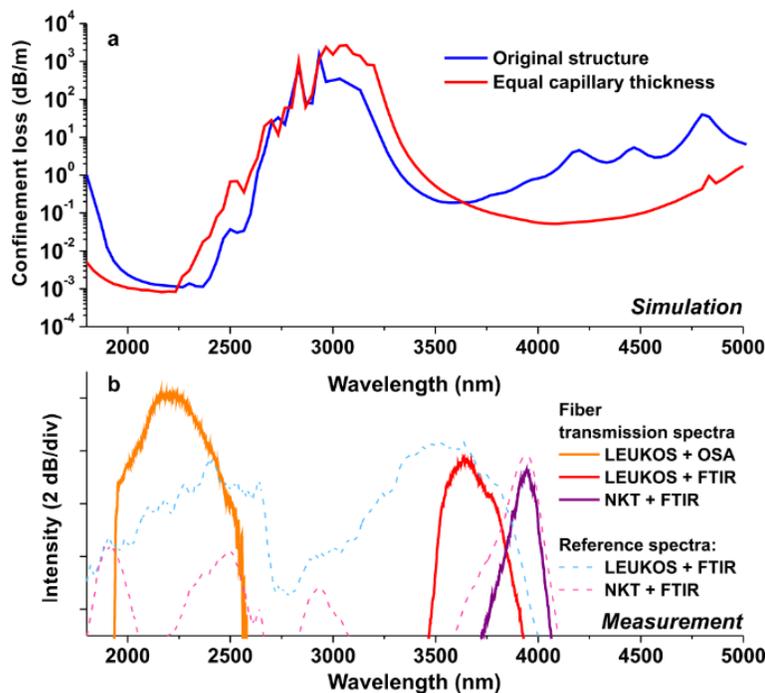

Fig. 3. (a) Finite element method simulation of the fiber transmission; (b) measured transmission spectrum for a 150 cm long fiber.

Attenuation of the fiber was measured by a standard cut-back method starting with a 12 m long sample, which was the longest uninterrupted length of the reported fiber available. The fiber was cut by 2 m in the following measurement steps, down to 4 m of final sample length. During the measurement, fiber was loosely looped over roughly 1 m of diameter, and no loops were made toward the end of measurement. A standard telecommunications fiber cleaver produced high-quality, repeatable cleaves and for every fiber length, the spectrum was recorded for three consecutive cleaves, to ensure measurement repeatability. Measured attenuation in the mid-infrared from around 3600 nm to 4000 nm was 1.5 dB/m and between 0.5 and 1.0 dB/m in the long-wave near-infrared, as shown in Fig. 4 (thick red traces). $LP_{01}$ mode losses obtained in simulations are up to three orders of magnitude lower around 2400 nm and a factor of two smaller around 4000 nm. This mismatch cannot be due to inclusion of silica material loss in modelling, because calculated power fraction in glass for this fiber is between $10^{-4}$ to $10^{-5}$. Surface scattering loss in the considered spectral range can account for 1 dB/km and does not explain the discrepancy, neither [3]. Measured attenuation level is qualitatively reproduced in simulations only when contribution of higher order modes is assumed, especially that of $LP_{02}$. Thus we relate attenuation of the fiber to content of higher order modes, which prompts for optimization of cladding and core diameters ratio in the discussed fiber. We note that dB/km straight-fiber loss has been reported in HCFs of the single capillary or "ice-cream cone" cladding designs in similar wavelength range [10,29].

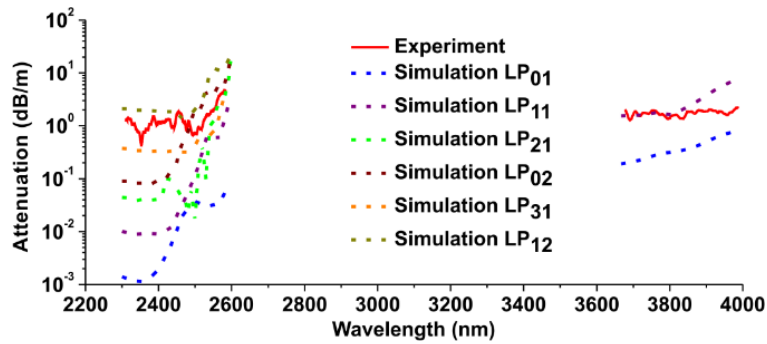

Fig. 4 Measured attenuation (solid traces) and calculated losses for different modes (dotted traces) of the developed NANF.

Bending loss measurement was performed by coiling of the 4 m long fiber sample into consecutive, full turns (full loops) of identical diameter, similarly to the procedure reported earlier for ARFs in [32]. Results – shown in Fig. 5 - were recorded for different bending radii from 40 mm down to 7.5 mm. In the mid-infrared window between 3500 nm and 3900 nm, the loss remains below 5 dB/m for bending radii of 15 mm or more, while for loops with radius of 30 mm or 40 mm, it remains below 2.5 dB/m down to 0.5 dB/m around 3800-3900 nm for 40 mm of bend radius. In the 1900-2600 nm window, a sharp short-wavelength transmission cut-off can be observed, which redshifts under bending with decreasing bending radius. This is consistent with earlier reported results in low ring number Kagomé ICFs, where redshift of short-wavelength bend loss edge has been assigned to bending-induced mode coupling between the core mode and leaky cladding modes [42] and has been numerically confirmed for NANFs, as well [5]. We note that in state-of-the-art single capillary-ring ARFs bending loss at the level of dB/km has been reported, although this related to the telecommunications wavelengths, i.e. less than 5 dB/km of loss (50 mm radius) over 1250-1650 nm [12] or over larger bending radii, i.e. 15 cm with 50 dB/km at around 3500 nm [10].

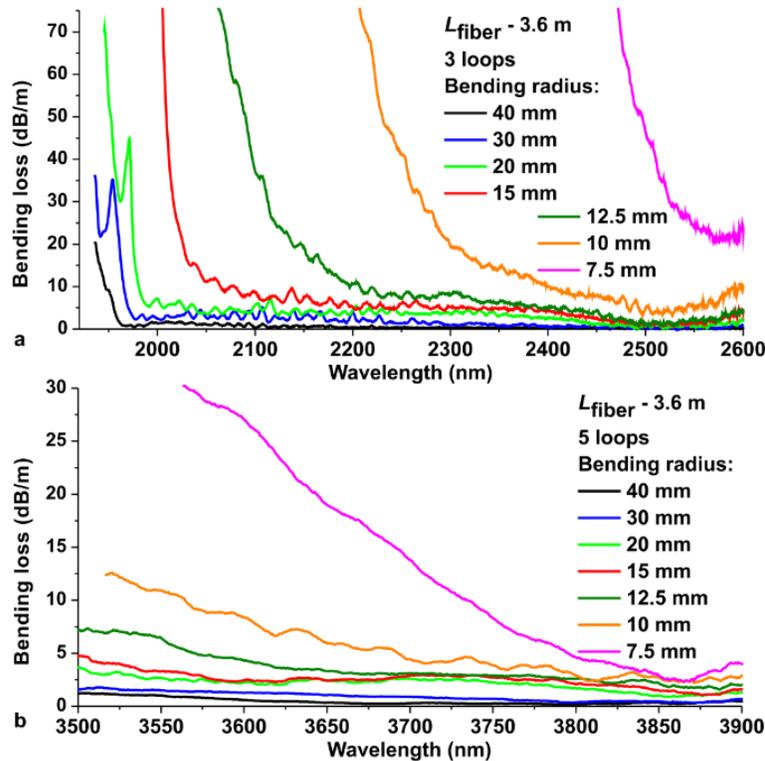

Fig. 5. Measured bend loss (a) over the long-wave infrared and (b) over the mid-infrared in the developed NANF.

In conclusion, we reported successful development of a silica nested capillary anti-resonant nodeless fiber with mid-infrared transmission up to around 4200 nm, while numerical simulations allow to anticipate this transmission to continue up to at least 5000 nm. This would cover frequencies of asymmetric stretching vibrational modes of $CO_2$ and $N_2O$ molecules (around 4250 nm and 4500 nm). The measured mid-infrared attenuation of 1-2 dB/m prompts for improvement of both the fiber design and the technological procedure. At the design level, the cladding capillary to the core diameter ratio should be optimized for effectively single-mode operation. Recent theoretical results reported for NANFs identify this ratio at around 1.129 [38], while in our case it is around 0.46. At the technological level, the pending improvement relates to maintaining of similar capillary wall thickness. Despite these imperfections, already with the demonstrated fiber were we able to achieve very low bending sensitivity corresponding to bending loss down to 0.5 dB/m at a bending radius of just 40 mm at 3800-3900 nm wavelengths, by introducing nested capillaries into the fiber microstructure. We note, that exceptional results in terms of attenuation and bending loss at a single dB/km level have been reported in hollow core conjoined tube capillary fibers and in NANFs, in each case in the third telecommunication window [12,15]. Tens of dB/km loss in effectively straight ARFs has been reported around 3000-4000 nm, e.g. [29,30], which also identified the challenge of bend loss in this wavelength range for bending radii smaller than 20 cm. Our results provide an experimental evidence for the feasibility of limiting bending loss to reasonable levels for bend radii as small as 3-4 cm and for the crucial role of the nested anti-resonant capillaries in providing high bend loss resistance in hollow-core silica fibers operating in the mid-infrared.

**Acknowledgements:** The authors acknowledge technical support of LEUKOS, Interlab, NKT Photonics, and Prof. R. Piramidowicz from Warsaw University of Technology.

**Funding.** Fundacja na rzecz Nauki Polskiej (FNP) (First TEAM/2016-1/1); Horizon 2020 Framework Programme (H2020) (722380); French Government and the French Embassy in Poland (2018); Région Bourgogne Franche-Comté (2018).

**References**
[1] F. Benabid, J. C. Knight, G. Antonopoulos, and P. S. J. Russell, "Stimulated Raman Scattering in Hydrogen-Filled Hollow-Core Photonic Crystal Fiber," Science 298(5592), 399-402 (2002).
[2] F. Couny, F. Benabid, P. J. Roberts, P. S. Light, and M. G. Raymer, "Generation and Photonic Guidance of Multi-Octave Optical-Frequency Combs," Science 318(5853), 1118–1121 (2007).
[3] A. D. Pryamikov, A. S. Biriukov, A. F. Kosolapov, V. G. Plotnichenko, S. L. Semjonov, and E. M. Dianov "Demonstration of a waveguide regime for a silica hollow - core microstructured optical fiber with a negative curvature of the core boundary in the spectral region > 3.5 μm," Opt. Express 19(2), 1441-1448 (2011).
[4] M. A. Duguay, Y. Kokubun, T. L. Koch, and Loren Pfeiffer, "Antiresonant reflecting optical waveguides in $SiO_2$-Si multilayer structures," Appl. Phys. Lett. 49(1), 13-15 (1986).
[5] F. Poletti, "Nested antiresonant nodeless hollow core fiber," Opt. Express 22(20), 23807–23828 (2014).
[6] C. Wei, R. Joseph Weiblen, C. R. Menyuk, and J. Hu, "Negative curvature fibers," Adv. Opt. Photonics 9(3), 504 (2017).
[7] P. Jaworski, F. Yu, R. R. J. Maier, W. J. Wadsworth, J. C. Knight, J. D. Shephard, and D. P. Hand, "Picosecond and nanosecond pulse delivery through a hollow-core Negative Curvature Fiber for micro-machining applications," Opt. Express 21(19), 22742-22753 (2013).
[8] X.-L. Liu, W. Ding, Y.-Y. Wang, S.-F. Gao, L. Cao, X. Feng, and P. Wang "Characterization of a liquid-filled nodeless anti-resonant fiber for biochemical sensing," Opt. Lett. 42(4), 863-866 (2017).
[9] U. Elu, M. Baudisch, H. Pires, F. Tani, M. H. Frosz, F. Köttig, A. Ermolov, P. ST.J. Russell, and J. Biegert, "High average power and single-cycle pulses from a mid-IR optical parametric chirped pulse amplifier," Optica 4(9), 1024-1029 (2017).
[10] B. Debord, A. Amsanpally, M. Chafer, A. Baz, M. Maurel, J. M. Blondy, E. Hugonnot, F. Scol, L. Vincetti, F. Gérôme, and F. Benabid, "Ultralow transmission loss in inhibited-coupling guiding hollow fibers," Optica 4(2), 209-217 (2017).
[11] M. Maurel, M. Chafer, A. Amsanpally, M. Adnan, F. Amrani, B. Debord, L. Vincetti, F. Gérôme, and F. Benabid, "Optimized inhibited-coupling Kagome fibers at Yb-Nd:Yag (85 dB/km) and Ti:Sa (30 dB/km) ranges," Opt. Lett. 43(7), 1598 (2018).
[12] S.-F. Gao, Y.-Y. Wang, W. Ding, D.-L. Jiang, S. Gu, X. Zhang, and P. Wang, "Hollow-core conjoined-tube negative-curvature fibre with ultralow loss," Nature Commun. 9:2828 (2018).
[13] M. N. Petrovich, F. Poletti, J. P. Wooler, A.M. Heidt, N.K. Baddela, Z. Li, D.R. Gray, R. Slavík, F. Parmigiani, N.V. Wheeler, J.R. Hayes, E. Numkam, L. Grűner-Nielsen, B. Pálsdóttir, R. Phelan, B. Kelly, John O'Carroll, M. Becker, N. MacSuibhne, J. Zhao, F.C. Garcia Gunning, A.D. Ellis, P. Petropoulos, S.U. Alam, and D.J. Richardson, "Demonstration of amplified data transmission at 2 μm in a low-loss wide bandwidth hollow core photonic bandgap fiber," Opt. Express 21(23), 28559-28569 (2013)
[14] J. R. Hayes, S. R. Sandoghchi, T. D. Bradley, Z. Liu, R. Slavík, M. A. Gouveia, N. V. Wheeler, G. Jasion, Y. Chen, E. N. Fokoua, M. N. Petrovich, D. J. Richardson, and F. Poletti, "Antiresonant Hollow Core Fiber With an Octave Spanning Bandwidth for Short Haul Data Communications," J. Lightwave Technol. 35(3), 437-442 (2017)
[15] T. D. Bradley, J. R. Hayes, Y. Chen, G. T. Jasion, S. R. Sandoghchi, R. Slavik, E. N. Fokoua, S. Bawn, H. Sakr, I.A. Davidson, A. Taranta, J. P. Thomas, M. N. Petrovich, D.J. Richardson, F. Poletti, "Record Low-Loss 1.3dB/km Data Transmitting Antiresonant Hollow Core Fibre,"2018 European Conference on Optical Communication (ECOC)
[16] X. Wang, D. Ge, Wei Ding, Yingying Wang, Shoufei Gao, Xin Zhang, Yizhi Sun, Juhao Li, Zhangyuan Chen, and Pu Wang, "Hollow-core conjoined-tube fiber for penalty-free data transmission under offset launch conditions," Opt. Lett. 44(9), 2145-2148 (2019).
[17] F. Benabid, F. Couny, J. C. Knight, T. A. Birks & P. St J. Russel, "Compact, stable and efficient all-fibre gas cells using hollow-core photonic crystal fibre," Nature 434, 488–491 (2005)
[18] P. T. Marty, J. Morel, and T. Feurer, "All-Fiber Multi-Purpose Gas Cells and Their Applications in Spectroscopy," Journal of Lightwave Technology 28(8), 1236-2340 (2010).

[19] F. Couny, F. Benabid, and P. S. Light, "Subwatt Threshold cw Raman Fiber-Gas Laser Based on H2-Filled Hollow-Core Photonic Crystal Fiber," Physical Review Letters 99, 143903 (2007).
[20] M. R. A. Hassan, F. Yu, W. J. Wadsworth, and J. C. Knight, "Cavity-based mid-IR fiber gas laser pumped by a diode laser," Optica 3(30), 218-221 (2016).
[21] F. B. A. Aghbolagh, V. Nampoothiri, B. Debord, F. Gerome, L. Vincetti, F. Benabid, and W. Rudolph, "Mid IR hollow core fiber gas laser emitting at 4.6 µm," Optics Letters 44(2), 383-386 (2019)
[22] M. Xu, F. Yu, and J. Knight, "Mid-infrared 1 W hollow-core fiber gas laser source," Opt. Lett. 42(20), 4055-4058 (2017)
[23] M. S. Astapovich, A. V. Gladyshev, M. M. Khudyakov, A. F. Kosolapov, M. E. Likhachev, and I. A. Bufetov, "Watt-level Nanosecond 4.42-µm Raman Laser Based on Silica Fiber," IEEE Photonics Technol. Lett. 31(1), 78-81 (2019)
[24] M. S. Habib, C. Markos, O. Bang, and M. Bache, "Soliton-plasma nonlinear dynamics in mid-IR gas-filled hollow-core fibers," Opt. Lett. 42(11), 2232–2235 (2017).
[25] M. S. Habib, C. Markos, J. E. Antonio-Lopez, and R. Amezcua-Correa, "Multioctave supercontinuum from visible to mid-infrared and bend effects on ultrafast nonlinear dynamics in gas-filled hollow-core fiber," Appl. Opt. 58(13), D7-D11 (2019).
[26] F. Köttig, D. Novoa, F. Tani, M. C. Günendi, M. Cassataro, J. C. Travers, and P. S. J. Russell, "Mid-infrared dispersive wave generation in gas-filled photonic crystal fibre by transient ionization-driven changes in dispersion," Nature Commun. 8, 813 (2017).
[27] A. I. Adamu, M. S. Habib, C. R. Petersen, J. Enrique Antonio Lopez, B. Zhou, A. Schülzgen, M. Bache, R. Amezcua-Correa, O. Bang, and C. Markos, "Deep-UV to Mid-IR Supercontinuum Generation driven by Mid-IR Ultrashort Pulses in a Gas-filled Hollow-core Fiber," Sci. Rep. 9, 4446 (2019).
[28] A. N. Kolyadin, A. F. Kosolapov, A. D. Pryamikov, A. S. Biriukov, V. G. Plotnichenko and E. M. Dianov, "Light transmission in negative curvature hollow core fiber in extremely high material loss region," Opt. Express 21(8), 9514-9519 (2013).
[29] F. Yu, W. J. Wadsworth, and J. C. Knight, "Low loss silica hollow core fibers for 3–4 µm spectral region," Opt. Express 20(10), 11153–11158 (2012).
[30] F. Yu and J. C. Knight, "Spectral attenuation limits of silica hollow core negative curvature fiber," Opt. Express 21(18), 21466 (2013).
[31] M. Michieletto, J. K. Lyngsø, C. Jakobsen, J. Lægsgaard, O. Bang, and T. T. Alkeskjold, "Hollow-core fibers for high power pulse delivery," Opt. Express 24(7), 7103 (2016).
[32] W. Belardi and J. C. Knight, "Hollow antiresonant fibers with low bending loss.," Opt. Express 22(8), 10091–10096 (2014).
[33] S.-F. Gao, Y.-Y. Wang, X.-L. Liu, W. Ding, and P. Wang, "Bending loss characterization in nodeless hollow-core anti-resonant fiber," Opt. Express 24(13), 14801–14811 (2016).
[34] R. M. Carter, F. Yu, W. J. Wadsworth, J. D. Shephard, T. Birks, J. C. Knight, and D. P. Hand, "Measurement of resonant bend loss in anti-resonant hollow core optical fiber," Opt. Express 25(17), 20612–20621 (2017).
[35] W. Belardi and J. C. Knight, "Hollow antiresonant fibers with reduced attenuation," Opt. Lett. 39(7), 1853-1856 (2014).
[36] W. Belardi, "Design and Properties of Hollow Antiresonant Fibers for the Visible and Near Infrared Spectral Range," Journal of Lightwave Technology 33(21), 4497–4503 (2015).
[37] A. F. Kosolapov, G. K. Alagashev, A. N. Kolyadin, A. D. Pryamikov, A. S. Biryukov, I. A. Bufetov, and E. M. Dianov, "Hollow-core revolver fibre with a double-capillary reflective cladding," Quantum Electron. 46(3), 267–270 (2016).
[38] Md. S. Habib, J. E. Antonio-Lopez, C. Markos, A. Schülzgen, and R. Amezcua-Correa, "Single-mode, low loss hollow-core anti-resonant fiber designs," Opt. Express 27(4), 3824-3836 (2019).
[39] J. E. Antonio-Lopez, S. Habib, A. V. Newkirk, G. Lopez-Galmiche, Z. S. Eznaveh, J. C. Alvarado-Zacarias, O. Bang, M. Bache, A. Schülzgen, and R. A. Correa, "Antiresonant hollow core fiber with seven nested capillaries," in 2016 IEEE Photonics Conference (IPC) (2016), pp. 402–403.
[40] M. S. Habib, O. Ba ng, and M. Bache, "Low-loss hollow-core silica fibers with adjacent nested anti-resonant tubes," Opt. Express 23, 17394–17406 (2015).
[41] H. Renner, "Bending losses of coated single-mode fibers: a simple approach," J. Light. Technol. 10(5), 544–551 (1992).
[42] M. Alharbi, T. Bradley, B. Debord, C. Fourcade-Dutin, D. Ghosh, L. Vincetti, F. Gérôme, and F. Benabid, "Hypocycloid-shaped hollow-core photonic crystal fiber Part II: Cladding effect on confinement and bend loss," Opt. Express 21(23), 28609-28616 (2013).